\documentclass[a4paper]{panl}
\usepackage{cite}
\usepackage{wrapfig}
\usepackage{graphicx}
\usepackage{amssymb}
\usepackage{amsfonts}
\usepackage{amsmath}
\usepackage{longtable}
\usepackage{rotating}
\usepackage{lscape}
\usepackage{epsfig}
\usepackage{multirow}
\usepackage{float}
\usepackage{slashed}

\newcommand{\forests}{\mathfrak{F}}
\newcommand{\infragr}{\mathfrak{I}'}

\originalTeX
\begin{document}

\title{A way of fast calculating lepton magnetic moments in quantum electrodynamics}
\maketitle
\authors{S.\,Volkov\/$^{a,b,}$\footnote{E-mail: sergey.volkov.1811@gmail.com, volkoff\_sergey@mail.ru}}
\setcounter{footnote}{0}
\from{$^{a}$\,Skobeltsyn Institute of Nuclear Physics at Lomonosov Moscow State University (Moscow, Russia)}
\from{$^{b}$\,Dzhelepov Laboratory of Nuclear Problems at Joint Institute of Nuclear Research (Dubna, Russia)}

\begin{abstract}
A new method of divergence subtraction in Feynman parametric integrals is presented. The method is suitable for calculating the lepton anomalous magnetic moments (AMM) in quantum electrodynamics (QED). The subtraction procedure eliminates all divergences before integration and leads to a finite Feynman parametric integral for each individual Feynman diagram. It is based on a forest formula with linear operators applied to the Feynman amplitudes of ultraviolet-divergent subdiagrams. The formula is similar to BPHZ; the difference is only in the linear operators used and in the way of combining them. The subtraction is equivalent to the on-shell renormalization from the beginning: for obtaining the final result we should only sum up the contributions of all Feynman diagrams after subtraction. The developed method is an improvement of the method presented by the author in 2016. The modification is specifically designed for calculating the contributions dependent on the relations of particle masses. In comparison with the old version, the new subtraction formula does not contain redundant terms and possesses some flexibility that can be used for improving the precision of calculations. Numerical test results are presented up to four loops.
\end{abstract}
\vspace*{6pt}

\section*{Introduction}\label{sec_intro}

The electron AMM $a_e$ is measured with a very high precision ~\cite{experiment} as well as the muon AMM $a_{\mu}$ ~\cite{muon_experiment}\footnote{the value is the statistical average of the new experimental value and the old one ~\cite{muon_experiment_old}}:
\begin{align}\label{eq_electron_experiment}
a_e[\text{expt.}]&=0.001\,159\,652\,180\,73(28), \\
\label{eq_muon_experiment}
a_{\mu}[\text{expt.}]&=0.001\,165\,920\,61(41).
\end{align}
The Standard Model prediction of $a_e$ uses the following representation:
{\small$$
a_e=a_e(\text{QED})+a_e(\text{hadronic})+a_e(\text{electroweak}), 
$$
$$
a_e(\text{QED})=\sum_{n\geq 1} \left(\frac{\alpha}{\pi}\right)^n
	a_e^{2n}, 
$$
$$
a_e^{2n}=A_1^{(2n)}+A_2^{(2n)}(m_e/m_{\mu})+A_2^{(2n)}(m_e/m_{\tau})+A_3^{(2n)}(m_e/m_{\mu},m_e/m_{\tau}),
$$}%
where $m_e,m_{\mu},m_{\tau}$ are the masses of the electron, muon and tau-lepton, respectively. Different terms of this expression were calculated by different researchers. A similar expression is used for $a_{\mu}$. The QED part forms the most significant contribution to both $a_e$ and $a_{\mu}$.

$A_1^{(2n)}$, $n=1,2,3,4$ have reliable double-checked values. However, $A_1^{(10)}$ is still sensitive in $a_e$ experiments, but is not double-checked yet. The value
\begin{equation}\label{eq_5loops_kinoshita}
A_1^{(10)}[\text{AHKN}]=6.737(159)
\end{equation}
was presented by T. Aoyama, M. Hayakawa, T. Kinoshita, M. Nio in 2019 ~\cite{kinoshita_atoms}. This value in combination with the remaining terms~\cite{kinoshita_atoms} and with the measured values of the fine-structure constant $\alpha$
\begin{align}\label{eq_alpha_rb2011}
\alpha^{-1}(\text{Rb-2011})&=137.035\,998\,996(85),
\\
\label{eq_alpha_cs2018}
\alpha^{-1}(\text{Cs-2018})&=137.035\,999\,046(27),
\\
\label{eq_alpha_rb2020}
\alpha^{-1}(\text{Rb-2020})&=137.035\,999\,206(11)
\end{align}
from ~\cite{alpha_rubidium,codata_2014}, ~\cite{alpha_cesium}, ~\cite{alpha_rubidium_2020} gives
{\small
$$
a_e[\text{theory},\text{Rb-2011},\text{AHKN}]=0.001\,159\,652\,182\,037(11)(12)(720), 
$$
$$
a_e[\text{theory},\text{Cs-2018},\text{AHKN}]=0.001\,159\,652\,181\,606(11)(12)(229), 
$$
$$
a_e[\text{theory},\text{Rb-2020},\text{AHKN}]=0.001\,159\,652\,180\,254(11)(12)(94),
$$}%
respectively. Each of these values has three uncertainties: from $A_1^{(10)}$ (statistical uncertainty of the Monte Carlo integration), the hadronic and electroweak corrections, the uncertainty of $\alpha$, respectively. Since the $\alpha$ uncertainty dominates, this calculation can be used for improving the precision of $\alpha$: (\ref{eq_electron_experiment}) in combination with (\ref{eq_5loops_kinoshita}) gives
$$
\alpha^{-1}[a_e,\text{AHKN}]=137.035\,999\,149\,6(331).
$$
The tension between it and (\ref{eq_alpha_rb2011}), (\ref{eq_alpha_cs2018}), (\ref{eq_alpha_rb2020}) amounts $1.69\sigma$, $2.43\sigma$, $1.61\sigma$.

In 2019 the author recalculated a part of $A_1^{(10)}$ and discovered a discrepancy with (\ref{eq_5loops_kinoshita}) ~\cite{volkov_5loops_prd}. The calculation gave the value
\begin{equation}\label{eq_5loops_volkov}
A_1^{(10)}[\text{Volkov+AHKN}]=5.862(90).
\end{equation}
The value leads to
{\small$$
a_e[\text{theory},\text{Rb-2011},\text{Volkov+AHKN}]=0.001\,159\,652\,181\,969(6)(12)(720), 
$$
$$
a_e[\text{theory},\text{Cs-2018},\text{Volkov+AHKN}]=0.001\,159\,652\,181\,547(6)(12)(229), 
$$
$$
a_e[\text{theory},\text{Rb-2020},\text{Volkov+AHKN}]=0.001\,159\,652\,180\,195(6)(12)(94),
$$
$$
\alpha^{-1}[a_e,\text{Volkov+AHKN}]=137.035\,999\,142\,7(331)
$$}%
with the same order of uncertainties. The last value has the tension $1.61\sigma$, $2.26\sigma$, $1.81\sigma$ with (\ref{eq_alpha_rb2011}), (\ref{eq_alpha_cs2018}), (\ref{eq_alpha_rb2020}), respectively. The shift is relatively small, but a significant error in $A_1^{(10)}$ would be sensitive in experiments.

A recent theoretical prediction for $a_{\mu}$ ~\cite{muon_sm_review} gives
$$
a_{\mu}=0.001\,165\,918\,095\,3(4386)(100)(10),
$$
where the uncertainties came from the hadronic, electroweak and QED contributions, respectively. It has the discrepancy $4.2\sigma$ with (\ref{eq_muon_experiment}). The QED uncertainty is relatively small, but it is based on the assumption that all the coefficient values are correct. $A_2$ and $A_3$ for the muon suffer from powered large logarithms of $(m_{\mu}/m_e)$. For example, in ~\cite{kinoshita_muon} the value
$$
A_2^{(10)}(m_{\mu}/m_e)=742.18(87)
$$
was published. It is more than 100 times larger than $A_1^{(10)}$ and is not double-checked yet. A significant error in it would give a shift in $a_{\mu}$ comparable with the hadronic uncertainty. Moreover, the higher-order terms are still important ~\cite{kinoshita_muon}. Thus, the development of a method of high-order QED calculcations is still actual. A possibility to reduce the influence of large logarithms in intermediate values is very important  for improving the precision.

\section*{Old and new method}

Let $G$ be a QED Feynman diagram contributing to the lepton AMM (it contains one external photon line and two external lepton lines). The old method ~\cite{volkov_2015} uses the linear operators $A$, $L$, $U$, where $A$ is the AMM projector multiplied by the Dirac matrix $\gamma_{\mu}$, $L$ is the standard on-shell renormalization operator for vertex-like diagrams, $U$ is an intermediate operator\footnote{See the definitions in ~\cite{volkov_2015}.}.

Let us examine the diagram from Fig. \ref{fig_example} contributing to $a_{\mu}$, where all lepton loops correspond to electrons. The old method gives the expression
{\small$$
\left[ A_G \left( 1-U_{G_e} \right)\left(1-U_{G_c}\right) - \left( L_G-U_G \right) A_{G_e} \left( 1-U_{G_c} \right) - \left( L_G-U_G \right) \left(1-L_{G_e}\right) A_{G_c} \right]
$$
$$ 
\times \left(1-U_{G_d}\right) \left(1-U_{c_1 c_2 c_3 c_4} \right) \left( 1-U_{c_1 c_2 c_3} - U_{c_1 c_3 c_4} \right) \left( 1-U_{a_1a_2} \right),
$$}%
where $G_c=aa_1a_2b_1b_2c_1c_2c_3c_4$, $G_e=aa_1a_2b_1b_2c_1c_2c_3c_4d_1d_2d_3e_1e_2e_3$, $G_d=aa_1a_2b_1b_2c_1c_2c_3c_4d_1d_2d_3$, the operator subscript means the subdiagram to which the operator is applied (encoded by the set of vertexes). Here all brackets should be expanded, each multiplication should be considered as a sequential transformation of the Feynman amplitudes from smaller to larger subdiagrams, all terms should be written in the same Feynman parametric space. In this example, the expression, that gives a contribution to $a_{\mu}$, contains $A$ applied to an electron subdiagram $G_c$. This looks like a nonsense, but the proof of the equivalence to the on-shell renormalization in ~\cite{volkov_2015} does not allow to avoid these nonsense terms.

\begin{figure}
	\begin{center}
		\includegraphics[width=50mm]{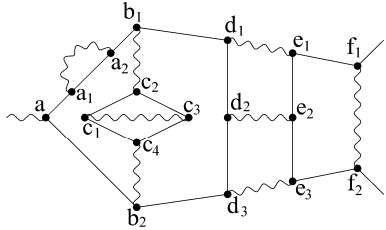}
		\vspace{-3mm}
		\caption{Example of a Feynman diagram for lepton anomalous magnetic moments.}
	\end{center}
	\labelf{fig_example}
	\vspace{-5mm}
\end{figure}

Also, an ability to preserve minimal gauge-invariant classes is important for divergence subtraction methods. Such a class is obtained from one diagram by moving internal photons along lepton loops and paths without jumping over the external photon (and without violating one-particle irreducibility). The old method preserves the classes without lepton loops ~\cite{volkov_5loops_prd}. However, the proof in ~\cite{volkov_2015} requires swapping the layers between subdiagrams to which $A$ is applied (like $G$, $G_c$, $G_e$); this can lead to exit from the class.

It surprisingly turned out that the drawbacks described above are fictitious: the nonsense terms are cancelled, and all gauge-invariant classes are preserved due to Ward identities of some kind. However, the old method remains redundand and nonflexible.

We use two types of Ward identities for individual lepton self-energy and vertex-like diagrams\footnote{See examples of applying Ward identities for individual diagrams in ~\cite{volkov_gpu}.}. The first type is $\Gamma_{\mu}(p,0)=-\partial\Sigma(p)/\partial p^{\mu}$, where $\Gamma_{\mu}(p,q)$ is a vertex-like Feynman amplitude, $\Sigma(p)$ is a lepton self-energy amplitude; the $\Gamma_{\mu}$-diagrams are obtained from the $\Sigma$-diagram by inserting the external photon into all possible places on the main lepton path\footnote{The main path is a lepton path between external lepton lines.}. The second type is $\Gamma_{\mu}(p,0)=0$; here the $\Gamma_{\mu}$-diagrams are obtained from some diagram by inserting the external photon into one lepton cycle. 

Let us describe the new method. We will use four operators $U_0,U_1,U_2,U_3$ instead of $U$. $U_0$ is applied to photon self-energy and photon-photon scattering subdiagrams and works as in the standard renormalization. $U_i,i=1,2,3$ are not fixed, but given by requirements (tentative):
\begin{itemize}
\item $U_i$ extracts the overall UV-divergent part completely;
\item $U_i$ preserves the Ward identity: if $\Gamma_{\mu}$ and $\Sigma$ (or $\Gamma_{\mu}$ itself) satisfy the Ward identity described above, then $U_i \Gamma_{\mu}$ and $U_i \Sigma$ (or $U_i \Gamma_{\mu}$ itself) satisfy it too;
\item $U_i$ cancels IR divergences of the subdiagram;
\item $U_i$ extracts the mass part completely: 
$$ U_i [a(p^2)+b(p^2) \slashed p]=a(m^2)+b(m^2)m+(\slashed p-m)(\ldots).$$
\end{itemize}
The definitions may differ for different particles. It is possible to put $U_3=L$.

By $\infragr[G]$ we denote the set of all vertex-like subgraphs of $G$ (including $G$) lying on the main path of $G$ and having the external photon of $G$. By $\forests[G]$ we denote the set of all forests of UV-divergent subdiagrams of $G$ containing $G$. The expression is
$$
\sum_{\substack{F=\{G_1,\ldots,G_n\}\in\forests[G] \\ G'\in \infragr[G] \cap F }} (-1)^{n-1} M^{G'}_{G_1} M^{G'}_{G_2} \ldots M^{G'}_{G_n}.
$$
Here $M^{G'}_{G''}$ equals $A_{G''}$, if $G''=G'$, $L_{G''}-(U_1)_{G''}$, if $G''=G\neq G'$, $L_{G''}$, if $G'\subset G''\subset G$,  $(U_0)_{G''}$, if $G''$ is a photon self-energy or a light-by-light subgraph. In the remaining cases it equals $(U_2)_{G''}$, if $G''$ lies on a lepton loop, $(U_w)_{G''}$, if $G''\in\infragr[G]$ and $G''\subset G'$, $(U_1)_{G''}$ in the other cases; here $w=3$, if $G$ has its external photon on a lepton loop, $w=1$ otherwise.

For the example from Fig.~\ref{fig_example} we have $\infragr[G]=\{G,G_e\}$ and the expression
$$
\left[ A_G \left( 1-(U_3)_{G_e} \right) - \left( L_G-(U_1)_G \right) A_{G_e} \right]\times \left(1-(U_2)_{G_c}\right)
$$
$$
\times \left(1-(U_0)_{G_d}\right) \left(1-(U_0)_{c_1 c_2 c_3 c_4} \right) \left( 1-(U_2)_{c_1 c_2 c_3} - (U_2)_{c_1 c_3 c_4} \right) \left( 1-(U_2)_{a_1a_2} \right). 
$$

\begin{figure}
	\begin{center}
		\includegraphics[width=0.9\textwidth]{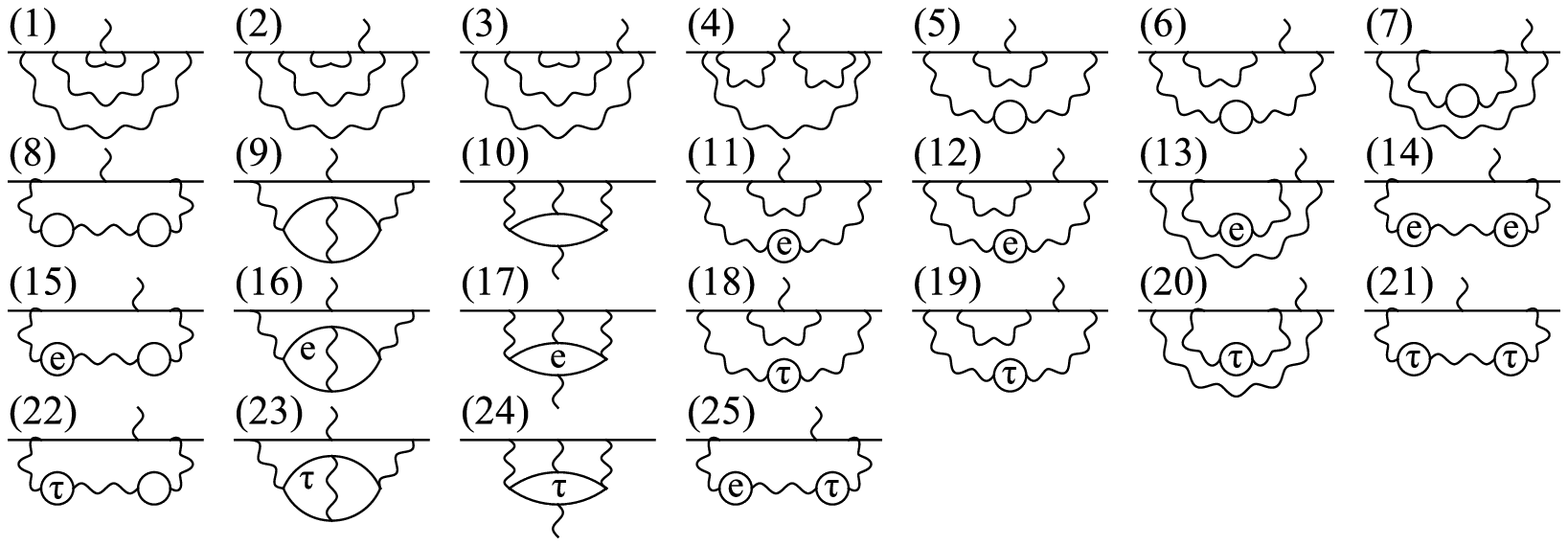}
		\caption{3-loop gauge-invariant classes for the muon $g-2$. Each class is given by one generating diagram. Electron and tau-lepton loops are marked with $e$ and $\tau$.}
	\end{center}
	\labelf{fig_3loops}
	%
	\begin{center}
		\includegraphics[width=0.5\textwidth]{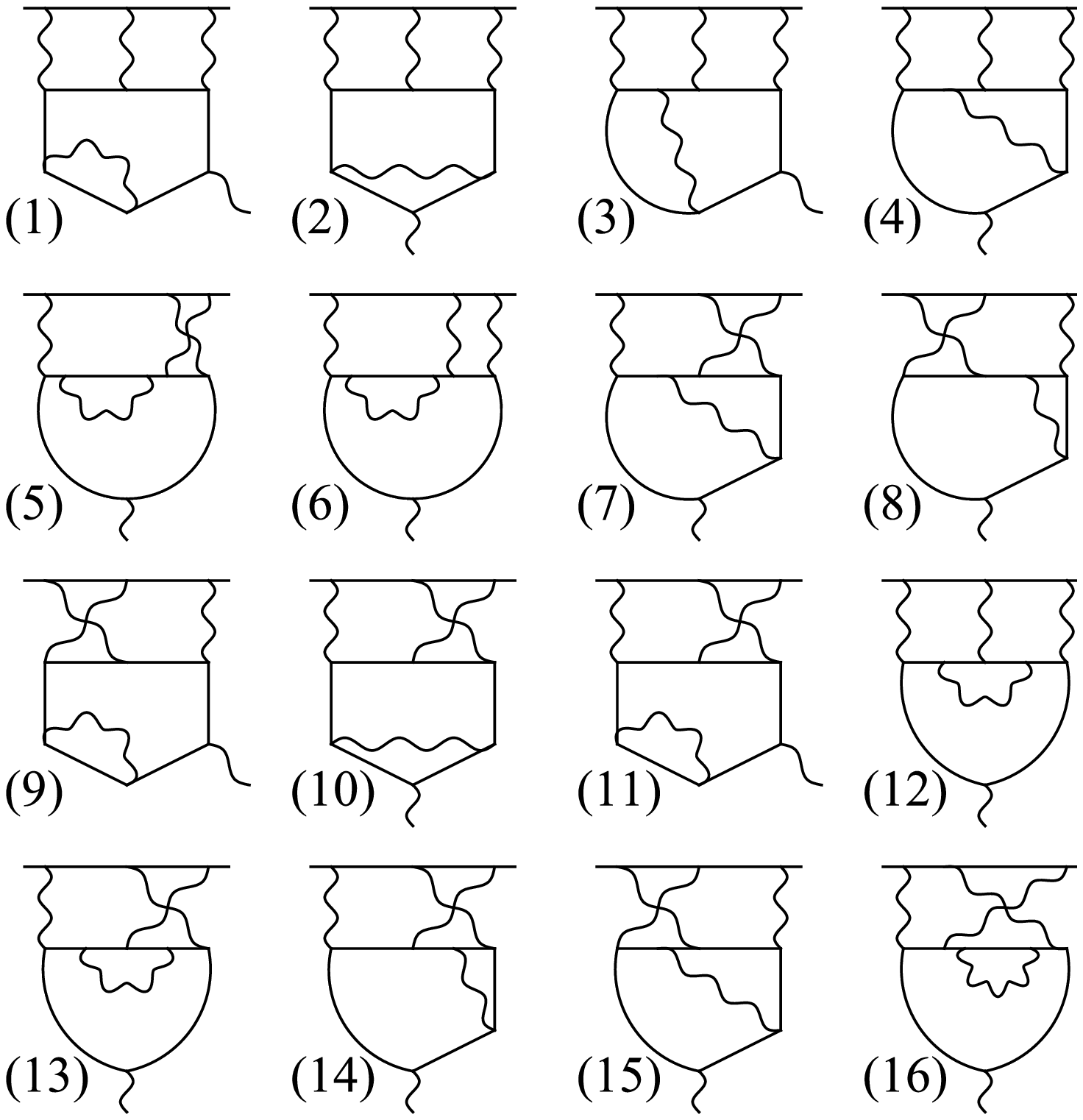}
		\hspace*{10mm}
		\includegraphics[width=0.385\textwidth]{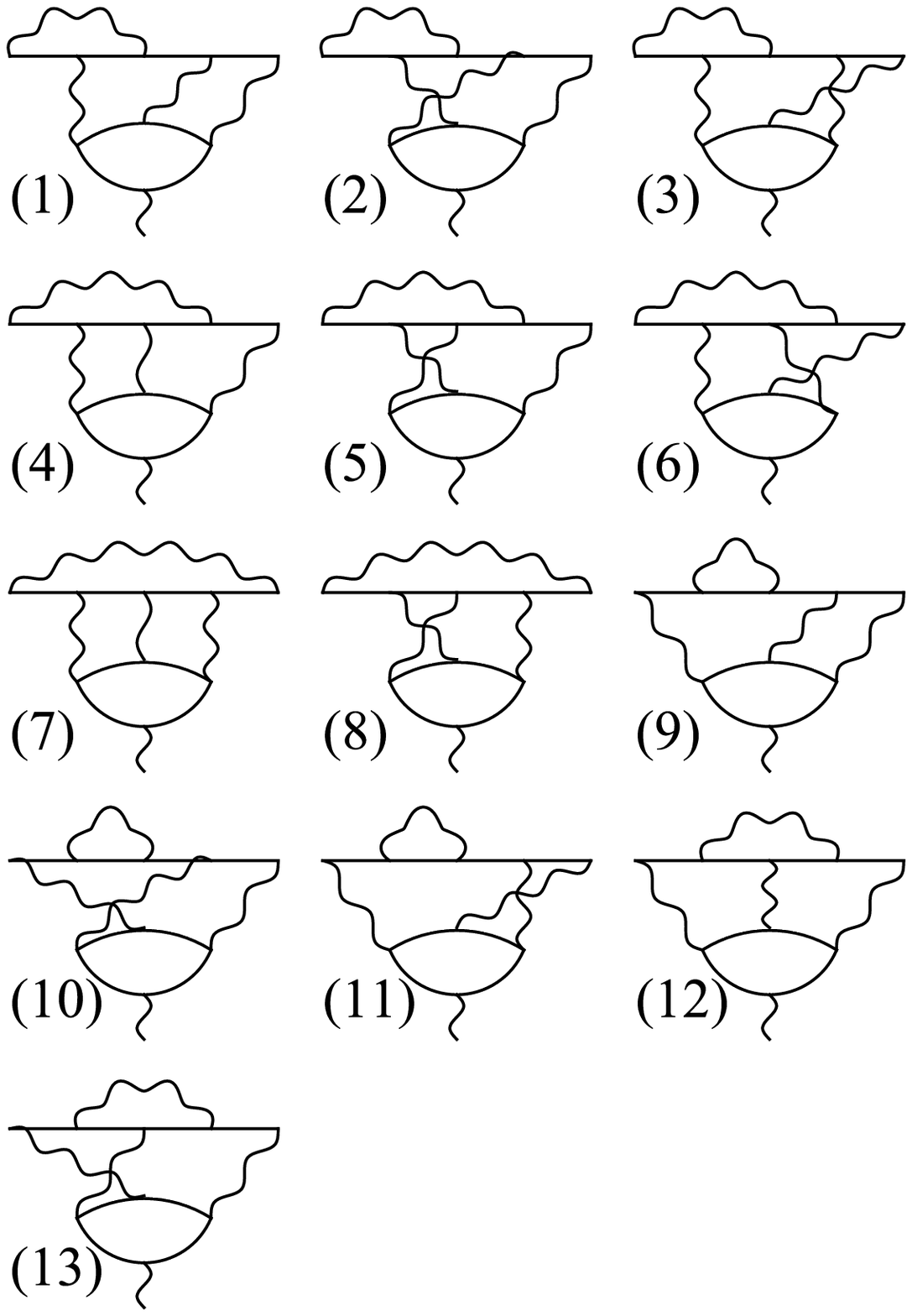}
		\caption{4-loop gauge-invariant classes IV(b) (left) and IV(c) (right) from ~\cite{kinoshita_muon}.}
	\end{center}
	\labelf{fig_IV_b_c}
	\vspace{-5mm}
\end{figure}

\begin{table}
\centering\footnotesize	
\caption{Contributions of the classes from Fig. \ref{fig_3loops} to $a_{\mu}^6$ and their comparison with known values obtained by different researchers with the help of analytical expressions.}\label{table_3loops} 

\medskip

\begin{tabular}{cccc}
\hline \hline 
Diagram & Our value & Analyt. value & Ref. \\ \hline 
1 & $0.448\,703(35)$ & $0.448\,70$ & ~\cite{analyt3,analyt_f,analyt_d,analyt_h} \\
2 & $-0.498\,224(67)$ & $-0.498\,25$ & ~\cite{analyt_e,analyt_c,analyt3,analyt_f,analyt_h,analyt_b}\\
3 & $0.533\,289(54)$ & $0.533\,36$ & ~\cite{analyt_b,analyt_e,analyt_d,analyt_c,analyt_f,analyt3,analyt_h}\\
4 & $0.421\,080(43)$ & $0.421\,17$ & ~\cite{analyt_b,analyt_e,analyt_c}\\ 
5 & $0.050\,178(16)$ & $0.050\,148\,7$ & ~\cite{analyt_b1,analyt_b4}\\
6 & $-0.112\,324(21)$ & $-0.112\,336$ & ~\cite{analyt_b1,analyt_b3}\\
7 & $-0.087\,987(12)$ & $-0.087\,984\,7$ & ~\cite{analyt_b1,analyt_b2}\\
8 & $0.002\,559\,8(15)$ & $0.002\,558\,5$ & ~\cite{analyt_mi}\\
9 & $0.052\,865(11)$ & $0.052\,87$ & ~\cite{analyt_mi}\\ 
10 & $0.370\,94(15)$ & $0.371\,005$ & ~\cite{analyt_ll}\\
11 & $1.617\,52(28)$ & - & - \\ 
12 & $-2.061\,83(39)$ & - & - \\ 
13 & $-1.948\,80(28)$ & - & - \\ 
11--13 & $-2.393\,11(56)$ & $-2.392\,391\,81(7)$ & ~\cite{analyt_mass_vp} \\
14 & $2.718\,85(62)$ & $2.718\,655\,7(2)$ & ~\cite{analyt_mass_vp} \\
15 & $0.100\,38(12)$ & $0.100\,519\,296(3)$ & ~\cite{analyt_mass_vp} \\
16 & $1.495\,45(92)$ & $1.493\,671\,80(4)$ & ~\cite{analyt_mass_vp} \\
17 & $20.947\,5(13)$ & $20.947\,1(29)$ & ~\cite{analyt_mass_ll} \\
18 & $0.000\,544\,10(20)$ & - & - \\
19 & $-0.001\,600\,08(30)$ & - & - \\ 
20 & $-0.001\,061\,37(17)$ & - & - \\
18--20 & $-0.002\,117\,35(40)$ & $-0.002\,117\,13$ & ~\cite{analyt_mass_vp} \\
21 & $0.000\,000\,276\,6(30)$ & $0.000\,000\,277\,833$ & ~\cite{analyt_mass_vp} \\
22 & $0.000\,038\,704(63)$ & $0.000\,038\,687\,5$ & ~\cite{analyt_mass_vp} \\
23 & $0.000\,295\,496(73)$ & $0.000\,295\,557$ & ~\cite{analyt_mass_vp} \\
24 & $0.002\,144\,3(12)$ & $0.002\,143\,31$ & ~\cite{analyt_mass_ll} \\
25 & $0.000\,524\,5(70)$ & $0.000\,527\,761$ & ~\cite{analyt_mass_3m}
\\ \hline \hline		
\end{tabular}
\end{table}


\begin{table}\centering
\footnotesize\addtolength\tabcolsep{-0.5mm}
\caption{Contributions of the diagrams from the class IV(b) with a tau-lepton loop (Fig. \ref{fig_IV_b_c}, left) to $a_{\mu}^8$ obtained by the old method ~\cite{volkov_2015} and by the new developed method ({New 2}).}\label{table_IV_b_mu_with_tau}	

\medskip

\begin{tabular}{cccccc}
\hline \hline Diagram & Old ~\cite{volkov_2015} & New 2 & Diagram & Old ~\cite{volkov_2015} & New 2 \\ \hline  
1 & $-0.045\,644(18)$ & $-0.045\,642(19)$ & 10 & $0.008\,448(22)$ & $0.016\,543(25)$ \\
2 & $-0.001\,059(16)$ & $-0.009\,095(15)$ & 11 & $0.016\,191(23)$ & $0.016\,173(25)$ \\
3 & $0.027\,752(20)$ & $0.027\,771(20)$ & 12 & $0.020\,202(21)$ & $0.020\,204(25)$ \\
4 & $-0.024\,477(15)$ & $-0.024\,472(15)$ & 13 & $0.002\,047(29)$ & $0.002\,011(32)$ \\
5 & $0.030\,956(24)$ & $0.030\,970(24)$ & 14 & $-0.011\,533(28)$ & $-0.011\,572(34)$\\
6 & $-0.014\,374(24)$ & $-0.014\,395(24)$ & 15 & $-0.004\,808(25)$ & $-0.004\,823(23)$  \\
7 & $0.005\,023(17)$ & $0.005\,006(17)$ & 16 & $0.018\,435(28)$ & $0.018\,450(30)$ \\
8 & $-0.041\,921(27)$ & $-0.041\,973(28)$ & $\sum $ & $ 0.006\,117(92)$ & $0.006\,058(96)$ \\
9 & $0.020\,880(24)$ & $0.020\,904(22)$ & & & 
\\ \hline \hline
\end{tabular}
\end{table}


\begin{table}
\centering\footnotesize
\caption{Contributions of the diagrams from the class IV(c) with an electron loop (Fig. \ref{fig_IV_b_c}, right) to $a_{\mu}^8$ obtained by the old method ~\cite{volkov_2015} and by the new developed method ({New 1}).}\label{table_IV_c_mu_with_e} 
			
\medskip
			
\begin{tabular}{cccccc}
\hline \hline Diagram & Old ~\cite{volkov_2015} & New 1 & Diagram & Old ~\cite{volkov_2015} & New 1 \\ \hline  
1 & $-118.181(23)$ & $-118.179(22)$ & 8 & $4.972(33)$ & $20.775(32)$ \\
2 & $91.815(22)$ & $91.841(21)$ & 9 & $100.498(24)$ & $100.475(24)$ \\
3 & $-80.851(20)$ & $-80.820(20)$ & 10 & $-76.896(32)$ & $-76.924(32)$ \\
4 & $2.607(21)$ & $2.620(22)$ & 11 & $89.117(31)$ & $89.140(31)$ \\
5 & $-74.660(27)$ & $-74.730(26)$ & 12 & $-26.755(13)$ & $-26.766(12)$ \\
6 & $37.012(23)$ & $37.016(22)$ & 13 & $-10.547(19)$ & $-10.543(21)$ \\
7 & $64.808(19)$ & $48.934(19)$ & $\sum$ & $2.940(87)$ & $2.840(86)$
\\ \hline \hline
\end{tabular}
\end{table}

Results of numerical tests of the methods are presented in Tables \ref{table_3loops}, \ref{table_IV_b_mu_with_tau}, and \ref{table_IV_c_mu_with_e}. We use two setups: \texttt{New 1} means $U_1=U_2=U$, $U_3=L$, \texttt{New 2} means $U_1=U_2=U_3=U$. We use fixed masses of particles without taking into account their uncertainty: $m_{\mu}=105.658\,374\,5\,\text{MeV}$, $m_e=0.510\,998\,95\,\text{MeV}$, $m_{\tau}=1\,776.82\,\text{MeV}$. All results from Table \ref{table_3loops} are in very good agreement with known values from literature. The difference between the methods used starts from four loops. The sum values in Table \ref{table_IV_b_mu_with_tau} are in good agreement with each other, with the value $0.006\,106(31)$ from ~\cite{kinoshita_muon}, and with ~\cite{kurz_4loops_mu_with_tau}. Diagrams 2 and 10 are the only diagrams with difference between the methods. The sum values in Table \ref{table_IV_c_mu_with_e} are in good agreement with the value $2.907\,2(44)$ from ~\cite{kinoshita_muon}. 
A difference between the methods is only in diagrams 7 and 8.

The author thanks Lidia Kalinovskaya, Gudrun Heinrich, Savely Karshenboim, Andrey Arbuzov for the important assistance, and Andrey Kataev for valuable consultations. Also, the author thanks the Laboratory of Information Technologies of JINR (Dubna, Russia) for providing an access to its computational resources and additionally the organizers of the conference FFK-2021 for providing a possibility to make a presentation. And beyond that, the author considers it proper to honor the memory of Fyodor Tkachov.

\bibliographystyle{pepan}
\bibliography{ffk_proceedings_2021_bib}

\end{document}